\documentclass[12pt,preprint]{aastex}

\catcode`\@=11
\newcommand{\gapprox}{\mathrel{\mathpalette\@versim>}}
\newcommand{\lapprox}{\mathrel{\mathpalette\@versim<}}
\newcommand{\propapprox}{\mathrel{\mathpalette\@versim\propto}}
\newcommand{\@versim}[2]
  {\lower3.1truept\vbox{\baselineskip0pt\lineskip0.5truept
\ialign{$\m@th#1\hfil##\hfil$\crcr#2\crcr\sim\crcr}}}
\catcode`\@=12

\shorttitle{EXPANSION OF YOUNGEST GALACTIC SNR G1.9+0.3}

\begin{document}

\title{Expansion of the  
Youngest Galactic Supernova Remnant G1.9+0.3}

\author{Ashley K. Carlton,\altaffilmark{1}
Kazimierz J. Borkowski,\altaffilmark{2}
Stephen P. Reynolds,\altaffilmark{2}
Una Hwang,\altaffilmark{3}
Robert Petre,\altaffilmark{3}
David A. Green,\altaffilmark{4}
Kalyani Krishnamurthy,\altaffilmark{5}
\& Rebecca Willett\altaffilmark{5}
}

\altaffiltext{1}{Department of Physics, Wake Forest University, Winston-Salem NC 27109; carlak7@wfu.edu} 
\altaffiltext{2} {Department of Physics, North Carolina State University,
  Raleigh NC 27695-8202}
\altaffiltext{3}{NASA/GSFC, Code 660, Greenbelt, MD 20771}
\altaffiltext{4}{Cavendish Laboratory; 19 J.J. Thomson Ave., 
Cambridge CB3 0HE, UK}
\altaffiltext{5}{Electrical and Computer Engineering, Duke University,
Durham NC 27708}

\begin{abstract}

We present a measurement of the expansion and brightening of G1.9+0.3,
the youngest Galactic supernova remnant, comparing {\sl Chandra} X-ray
images obtained in 2007 and 2009.  A simple uniform expansion model
describes the data well, giving an expansion rate of $0.642 \pm
0.049$\% yr$^{-1}$, and a flux increase of $1.7 \pm 1.0$\% yr$^{-1}$.
Without deceleration, the remnant age would then be $156 \pm 11$ yr,
consistent with earlier results.  Since deceleration must have
occurred, this age is an upper limit; we estimate an age of about 110
yr, or an explosion date of about 1900.  The flux increase is
comparable to reported increases at radio wavelengths.  G1.9+0.3 is
the only Galactic supernova remnant increasing in flux, with
implications for the physics of electron acceleration in shock waves.

\end{abstract}


\section{Introduction}
\label{intro}
The small Galactic supernova remnant (SNR) G1.9+0.3 was determined to
have an age of less than $140 \pm 30$ yr based on a comparison of a
radio image from 1985 and an X-ray image from 2007 (Reynolds et
al.~2008; Paper I).  This makes G1.9+0.3 the remnant of the most
recent known Galactic supernova.  Its X-ray spectrum is highly
absorbed ($N_H \sim 5 \times 10^{22}$ cm$^{-2}$; Reynolds et al.~2009
[Paper III]) and dominated by synchrotron emission with a very high
rolloff energy, $h\nu_{\rm roll} \sim 2.2$ keV.  Faint thermal
emission can be isolated from the N rim, with line emission from Si,
S, Ar, Ca, and Fe, as well as a 4.1 keV line from $^{44}$Sc produced
in the decay chain from $^{44}$Ti (Borkowski et al.~2010; Paper IV).

The mean expansion of G1.9+0.3 between 1985 and 2007 was estimated to
be $16 \pm 3$\%, or $(0.73 \pm 0.14)$\% yr$^{-1}$ (Paper I).  A
comparison of the 1985 radio image to a new 2008 VLA\footnote{The VLA
is part of the National Radio Astronomy Observatory, a facility of the
National Science Foundation operated under cooperative agreement by
Associated Universities, Inc.}  radio image confirmed this result,
giving a mean expansion rate of 0.65\% yr$^{-1}$ (Green et al.~2008;
Paper II).  At an assumed distance of 8.5 kpc, consistent with the
high X-ray absorption, the implied shock speed is about 14,000 km
s$^{-1}$, comparable to that inferred from measured line widths (Paper
IV).  The high velocity, and the symmetric nonthermal X-ray
morphology, suggest a Type Ia origin, though this is far from certain.

We observed G1.9+0.3 with {\sl Chandra} again in 2009 for a total of
237 ks, and now have the opportunity of examining the expansion
between two X-ray observations.  Since the radio and X-ray
morphologies differ somewhat (Paper I), it is important to determine
an expansion from X-ray data alone, which should give the best
indication of the blast wave speed.

\section{Observations}
\label{obssec}

Our 2007 observation with $Chandra$ using the ACIS-S CCD camera (S3
chip) took place on 2007 February 10 for 24 ks, and March 4 for 26 ks.
We checked aspect correction and created new level 1 event files
appropriate for VFAINT mode.  No flares occurred during the
observation.  CTI correction was applied and calibration was performed
using CALDB version 4.3.1.  Finally, the data sets were merged and
weighted response files created.  About 8000 counts were obtained.  We
reobserved on 2009 July 13 and July 27 for a total of 237 ks.  The
same procedures were followed.  About 40,000 counts were obtained.  We
verified that the data sets were well aligned for both observations by
using the fitting method described below to compare pairs of
sub-observations of the same epoch.  We found negligible shifts (far
smaller than a pixel).  The weighted average interval between
observations was 2.408 yr.

We used an elliptical mask, chosen by inspection, to exclude
non-source parts of the image, allowing the analysis to proceed
faster.  The mask also excluded circular regions surrounding several
point sources.  We checked that the mask did not skew the results by
repeating the analysis without it; the results did not change
significantly.  Therefore, all results below were obtained with the
mask.  In addition, we subtracted a constant background, estimated
from a region of the 2009 image free of discrete sources.

\section{Method}

To measure the expansion, we compared the 2009 platelet-smoothed
(Willett 2007) image to the raw 2007 data.  (We tested for dependence
on smoothing method by repeating the analysis with an image smoothed
using the spectro-spatial method of Krishnamurthy, Raginsky, \&
Willett (2010).  The results were not significantly different, either
in best-fit values or in uncertainties.)  The model contains four
parameters: a physical scaling factor, a surface brightness scaling
factor, and expansion center coordinates.  We shrank the
background-subtracted image by an overall multiplicative factor about
a point chosen by eye to be near the center, scaled its brightness
(counts/shrunken pixel) by another factor, added back a background
taken from the 2007 image, and shifted the image as necessary to
minimize our fit statistic described below.  The shift from the
originally chosen origin produced an ``expansion center,'' just the
point on the 2009 image from which this simple uniform-expansion model
yields the closest approximation to the 2007 data.  This method is
analogous to the one used by \citet{vink08} for Kepler's supernova
remnant.

Because the number of counts per bin is small, the 2007 data are not
well-represented by a Gaussian distribution, making the use of the
$\chi^2$ statistic inappropriate.  The data are better described by a
Poisson distribution: the probability of getting a
particular set of data counts $\{D_i\}$ is given by
\begin{equation}\label{Likelihood}
L = \prod_{i }^{ } \frac{ { {M }_{i } }^{ {D }_{i } } }{ {D }_{i }!
}\exp(- {M }_{i } ) \; ,
\end{equation}
where $M_i =S_i +B_i$ is the sum of source and background model
expectation values.  The maximum-likelihood fit can be found by minimizing the
Cash statistic \citep{cash79} 
given by
\begin{equation}\label{C-statistic}
C = 2 \sum_{i} \left[ {M }_{i }- {D }_{i }+ {D }_{i } \left ( \log D_i
-\log{M }_{i} \right ) \right ] ^{ }
\end{equation}
In the limit of large numbers of counts, the $C$-statistic becomes
identically $\chi^2$.  The minimum was found using a downhill simplex
algorithm. The resulting values for the parameters were found to be
independent of the choice of initial values.

The method produces a factor that represents a change in the surface
brightness (counts/pixel) (after scaling by the ratio of exposure
times, a factor of 4.779).  However, the effective change in pixel
size as the 2009 model is shrunk to match the 2007 data means that the
X-ray flux changes by an additional factor, just the linear scaling
factor squared.  Our best-fit value for the linear scale change is
$(1.55 \pm 0.12)$\%, or an expansion rate of $(0.642 \pm 0.049)$\%
yr$^{-1}$.  Our brightness scaling factor by which the 2009 model had
to be diminished, before correcting for the exposure-time difference,
was $0.2064 \pm 0.0042$, indicating a surface-brightness change of a
factor of $1.01 \pm 0.02$ over the 2.408-year interval.  After
correction for the size increase, this yields a mean flux
increase of a factor of $(4.55 \pm 2.08)$\% or $(1.89 \pm 0.86)$\%
yr$^{-1}$.  The derived expansion center is shown as the cross on the
X-ray image of Fig.~1.  Our results are summarized in Table 1.
All errors are 90\% confidence intervals.

We also performed Markov chain Monte Carlo (MCMC) simulations using
the PyMC software package \citep{patil10}.  These simulations
converged on the same optimal values of the parameters found using our
optimization routine presented in Table 1, to an accuracy much better
than our reported errors, supporting the validity of the
maximum-likelihood method.

By inspection, the remnant appears to have spatial variations in its
relative expansion factor and change in brightness.  The east and west
X-ray bright lobes appeared to expand by a larger factor than the
radio-bright northern region and the dimmer southern region.  To see
if this visual impression was supported statistically, we divided the
remnant into four regions and performed a joint fit allowing the
regions to have separate expansion and brightness factors, but sharing
a common expansion center.  The resulting fit was not a significant
improvement on the four-parameter fit.  Confirmation of spatial
variations in the expansion rate will require a deeper observation.

\section{Flux Increase}

The flux increase can be measured directly by performing a joint
analysis of spatially-integrated spectra.  We used a simple absorbed
power-law model, but instead of subtracting background we modeled it
as a combination of sky and particle components (Paper III).
Background spectra were extracted from an annulus surrounding the
remnant (Paper III, Fig.~2). Source and background spectra from both
epochs were fit jointly, using MCMC simulations as implemented in the
X-ray software package XSPEC \citep{arnaud96}. We assumed flat priors
for the absorbing column density $N_H$ and the power law spectral
index $\Gamma$, and a logarithmic prior for (absorbed) 2007 and 2009
X-ray fluxes in the energy range from 1 to 7 keV. The fitted
absorption $N_H$ is $7.23 (7.04, 7.42) \times 10^{22}$ cm$^{-2}$,
$\Gamma = 2.40 (2.33, 2.46)$, $F_{1-7~{\rm keV}}({\rm 2007}) = 2.70
(2.64, 2.76) \times 10^{-12}$ ergs cm$^{-2}$ s$^{-1}$, and
$F_{1-7~{\rm keV}}({\rm 2009}) = 2.81 (2.78, 2.84) \times 10^{-12}$
ergs cm$^{-2}$ s$^{-1}$.  (We neglect dust scattering, so the value of
$N_H$ here includes contributions from both absorption and
scattering.)  The best estimate of the flux increase is 4.1\%, but
uncertainties are large (the 90\%\ confidence interval is
(1.7\%-6.6\%), using the 0.05 and 0.95 quantiles of the MCMC draws).
Since the thermal contribution to the integrated spectrum is small,
the increase is evidently in the synchrotron emission.

The rate of X-ray flux increase is $1.7 \pm 1.0$\% yr$^{-1}$. This is
consistent with our image-based result, and comparable to the rate of
increase in the radio flux density, estimated at 2\%\ yr$^{-1}$ in
Paper II and $1.22^{+0.24}_{-0.16}$\%\ yr$^{-1}$ by \citet{murphy08}
(1$\sigma$ errors).

\section{Discussion}

For a distance of 8.5 kpc and a mean shock radius of about $50''$ (the
radius of the bright ring in Fig.~1), our measured expansion rate
gives a shock velocity of 13,000 km s$^{-1}$; the SE-NW extensions
beyond the bright ring are at about $60''$ for a shock velocity of
almost 16,000 km s$^{-1}$.  These values bracket the spectroscopically
deduced velocities of order 14,000 km s$^{-1}$ (Paper IV) and
indicate that our distance estimate is a good approximation.

While our measured expansion rate gives an upper limit to the age of
G1.9+0.3 of 156 yr, it is likely considerably younger because of
blast-wave deceleration.  Significant deceleration is expected for
either a Type Ia or core-collapse (CC) SN explosion.  The simplest
Type Ia SNR model consists of ejecta with an exponential ejecta
density profile, $\rho_e \propto \exp(-v/v_e)t^{-3}$, expanding into a
uniform ambient ISM \citep{dwarkadas98}. The velocity scale $v_e$ is
equal to $2440 E_{51}^{1/2} \left(M_e/M_{Ch}\right)^{-1/2}$ km
s$^{-1}$, where $E_{51}$ and $M_e/M_{Ch}$ are ejecta kinetic energy
(in $10^{51}$ ergs) and mass (in units of the Chandrasekhar mass
$M_{Ch}$).  The free expansion velocity of ejecta at the reverse shock
is very high, $v=19600 \left(R_r/2~{\rm pc}\right) \left( t/100~{\rm
yr} \right)^{-1}$ km s$^{-1}$, or $v/v_e = 8.03 \left(R_r/2~{\rm
pc}\right) \left( t/100~{\rm yr} \right)^{-1} E_{51}^{-1/2}
\left(M_e/M_{Ch}\right)^{1/2}$, where $R_r$ and $t$ are the reverse
shock radius and age of G1.9+0.3. At very early times, the separation
between forward and reverse shocks is a small fraction of their radii,
$\sim 10\%$ in the model of \citet{dwarkadas98} with the standard
postshock compression $r$ of 4 (corresponding to the adiabatic index
$\gamma=5/3$), and less for a cosmic ray modified blast wave with a
larger compression. Using mass conservation \citep{mckee74}, this
fractional separation can be estimated as $\sim (r/(r-1))^{1/3}-1$.
There is growing evidence for $r>4$ in young SNR; e.g.,
\citet{williams11} found a large ($r \sim 12$) blast wave compression
in the LMC SNR 0519$-$69.0.  In their hydrodynamical model with
$\gamma = 1.18$, the shocked region thickness is only $\sim 3\%$ at
early times.  We then used a thin-shell approximation ($R_r \cong R$);
for ambient medium mass $M$ varying as $r^\beta$, eqs.~(37) and (38)
in \citet{chevalier05} reduce to
$\left([\beta+1]y-1-4x-6x^2\right)\left(x^2d\,y/d\,x+[1-x]y\right) =
\beta y \left(1+3x[1+2x+2x^2]\right)$, where
$y=\frac{2}{\beta+1}\frac{M}{M_e}\frac{v_e}{v} \exp(v/v_e)$ and
$x=\frac{v_e}{v}$.  At very early times, $y \rightarrow 1$ for
well-behaved, asymptotic thin-shell solutions.  This imposes an
initial condition $y=1$ at $x=0$; it is straightforward to verify that
$y=1+4x+6x^2$ is an exact solution of the ordinary
differential equation for $y$ that satisfies this initial
condition. For uniform ambient medium ($\beta=3$), we arrive at an
analytic, parametric solution for the dimensionless blast wave radius
$r' \equiv R/R' = \left(1+4v_e/v+6(v_e/v)^2\right)^{1/3} \left(2v/v_e
\right)^{1/3}\exp(-v/3v_e)$, where $R'=2.19
\left(M_e/M_{Ch}\right)^{1/3}n_0^{-1/3}$ pc and $n_0$ is preshock H
density in cm$^{-3}$.  A good match to the measured expansion rate of
$0.64\%$ yr$^{-1}$ and the remnant's radius of 2 pc is provided by a
fiducial SN model with $E_{51}=1$ and $M_e=M_{Ch}$ at an age of 110 yr
and with $n_0=0.022$ cm$^{-3}$. For this model, $v/v_e=7.3$,
$r'=0.25$, the dimensionless time $t' \equiv
r'\left(v/3^{1/2}2v_e\right)^{-1} = 0.12 $, and the deceleration
parameter $m \equiv d \ln R /d \ln t = 0.69$.

In the CC scenario, we favor explosions of stripped compact cores over
red supergiant explosions, because of the very high shock velocity and
the presence of shocked Fe and other heavy elements in the outer
ejecta layers.  The outermost ejecta in explosions of stripped massive
cores (WR stars) are well described by a power law density profile,
$\rho_e \propto v^{-n}t^{-3}$ with $n=10.2$ \citep{matzner99}.  The
blast wave radius $r$ initially increases as $t^m=t^{(n-3)/(n-2)}$ as
SN ejecta are expanding into a WR stellar wind \citep{chevalier82}, so
that $m = 0.88$. \citet{chevalier06} find that $m$ drops slightly to
0.866 $\sim 20$ yr after the explosion (for a representative WR
progenitor), corresponding to $n=9.5$ (the ejecta density profile
becomes less steep as $v$ decreases). Since X-ray and radio emission
decreases with time while the blast wave propagates through the wind,
G1.9+0.3 cannot be at this stage of evolution. Instead, the blast wave
must have passed the WR termination shock and now be propagating into
the shocked wind with an approximately constant density.  The WR
termination shock can be less than 2 pc close to the Galactic center
where the ISM pressure is much higher than in the solar
vicinity. \citet{chevalier04} investigated the evolution of bubbles
blown by WR stars in high-pressure environments, and found that the
wind termination shock $R_t$ stalls at $R_t \approx 1.8 v_8^{1/2}
\dot{M}_{-5}^{1/2} \left(p/k/10^6~{\rm cm}^{-3}K\right)^{-1/2}$ pc for
ISM pressures $p/k$ higher than $2.5 \times 10^5
\dot{M}_{-5}v_8\left(t/t_{WR}\right)^{-2}$ cm$^{-3}$ K ($v_8$,
$\dot{M}_{-5}$, and $t$ are the WR wind speed, mass loss rate, and
duration in units of 1000 km s$^{-1}$, $10^{-5} M_\odot$ yr$^{-1}$,
and $t_{WR}= 3 \times 10^4$ yr, respectively).  If the current blast
wave radius is much larger than the wind termination shock, then $m =
(n-3)/n$ (or 0.68 for $n=9.5$; Chevalier 1982), comparable to our
estimates for the Type Ia scenario.  Intermediate values between 0.68
to 0.87 arise if the termination shock radius is a sizable fraction of
the current remnant radius.  This will affect the estimated age. With
$m=0.7$ (0.8), G1.9+0.3 is 110 (125) yr old.  Ksenofontov et
al.~(2010) modeled G1.9+0.3 with power-law ejecta with $n = 7$,
obtaining an age of only 80 yr (they also relied on estimates of the
remnant's expansion from Papers I and II).
 
Realistic high-velocity ejecta density profiles in Type Ia SNe may
deviate substantially from the idealized exponential profile. High
velocity features are frequently seen in Type Ia SN spectra at early
times, mostly in \ion{Ca}{2} absorption, suggesting the presence of
density structures of unknown origin.  Type Ia explosion models differ
significantly in predicted density profiles at high velocities,
including variations in steepness and the presence or absence of
distinct density structures.  This strongly affects the deceleration
of the blast wave, as demonstrated by \citet[][Figs.~2 and
3e]{badenes03}. It is then highly desirable to measure the
deceleration directly with present and future {\it Chandra}
observations. This may be feasible with observations spanning the
anticipated {\it Chandra} lifetime (we hope at least one more
decade). Assuming no deceleration, the current expansion age of 156 yr
at the mean epoch 2008 would increase to 166 yr at epoch 2018;
alternatively, the expansion rate would decrease to $0.603$\% yr
$^{-1}$ from the current value of $0.642$\% yr$^{-1}$. With $m = 0.7$,
we expect a smaller rate of $0.588$\% yr$^{-1}$ in 2018. It may be
possible to measure this difference, since significant improvements
are expected with upcoming long {\it Chandra} observations, on account
of much better photon statistics and the increased time baseline.

The flux increase we report of about 1.7\% yr$^{-1}$ is consistent
with both our estimate of the rate of increase of radio flux (about
2\%; Paper II), and that reported by Murphy et al.~(2008),
$1.22^{+0.24}_{-0.16}$\% yr$^{-1}$.  As discussed in Paper II, radio
flux increases are difficult but not impossible to produce; in the
absence of increasing efficiency of magnetic-field amplification
and/or particle acceleration, they typically require expansion into a
uniform medium.  We should note that prompt radio emission from
supernovae is observed to drop rapidly with time at frequencies at
which it is optically thin \citep[e.g.,][]{panagia06}; this emission
is typically modeled as synchrotron emission from a region occupied by
magnetic field and relativistic electrons, with a constant fraction of
shock energy going into each component, as a shock moves through a
dense stellar wind.  The increasing radio and X-ray flux from G1.9+0.3
argues against current interaction with a stellar wind. In the
formalism of Katsuda et al.~(2010), Appendix A, for expansion into a
uniform medium, the intensity at radio wavelengths $I_\nu$ varies at
the fractional rate $I_\nu^{-1} dI_\nu/dt = p/t$, where $p = m(4 +
\alpha) - (3 + \alpha)$ and $I_\nu \propto \nu^{-\alpha}$.  In that
work, the relativistic-electron density and magnetic-field energy
density are both assumed proportional to the post-shock energy density
$\rho u_{\rm sh}^2$.  Then for $\alpha = 0.62$, appropriate for the
integrated radio spectrum of G1.9+0.2, we have $p = 4.62m - 3.62.$ The
integrated flux $S_\nu \propto I_\nu R^2$, giving a fractional rate
${\cal R} \equiv S_\nu^{-1} dS_\nu/dt = (p + 2m)/t$.  For G1.9+0.3, we
consider two cases: undecelerated expansion ($m = 1$ and $t \sim 156$
yr), and $m = 0.7$ ($t \sim 110$ yr).  The undecelerated case then
gives a fractional flux increase rate of about 2\% yr$^{-1}$, while
the decelerated case gives 1\% yr$^{-1}$, so that the radio results
nicely bracket the theoretical expectations.

For a power-law electron distribution with an exponential cutoff above
some $E_{\rm max}$, the rate of change of intensity and flux will be
different in general for observing frequencies $\nu > \nu_c \equiv
1.82 \times 10^{18} E_{\rm max}^2 B$, the characteristic synchrotron
frequency emitted by electrons with energy $E_{\rm max}$ in a magnetic
field $B$ (e.g., Pacholczyk 1970; cgs units).  If $\nu_c$ rises with
time, the X-ray flux will rise faster (or decline more slowly) than
the radio, while if $\nu_c$ drops with time, the reverse is true.  So
the question of whether the X-ray rate of change is larger or smaller
than the radio is of considerable theoretical interest.  The same
Appendix from Katsuda et al.~(2010) gives the additional fractional
rate of change of intensity (or flux) as $\Delta{\cal R} =
(a/2t)\sqrt{\nu/\nu_c}$, where $\nu_c \propto t^a$.  For an average
over the spectrum of G1.9+0.3, we take $\nu/\nu_c = 1.5$ (or $h\nu =
3.3$ keV, about the peak in the integrated spectrum).  The behavior
with time of $\nu_c$ depends on the mechanism limiting electron
acceleration.  If radiative losses limit it, Katsuda et al.~(2010)
show that $a = 2m - 2$, while if the finite remnant age is the
limitation, $a = 7(m - 1) + 2$.  Then undecelerated motion predicts
the same rate of change in X-rays as radio for loss-limited
acceleration, and a faster rate, about 2.8\% yr$^{-1}$, for
age-limited acceleration.  For decelerated expansion, both
possibilities give slower rates of increase: about 0.6\% yr$^{-1}$ for
loss-limited, but only slightly less than 1\% yr$^{-1}$ for
age-limited acceleration.  Subsequent observations should allow both
the radio and X-ray rates of flux increase to be determined more
accurately, offering the possibility of determining the mechanisms
limiting particle acceleration in G1.9+0.3.  We point out that
age-limited acceleration would have to limit ion acceleration as well,
while if electrons are loss-limited, ions might well be accelerated to
much higher energies than those we deduce from electron synchrotron
emission.

\section{Conclusions}

\begin{enumerate}
\item The change in morphology of the X-ray emission from G1.9+0.3 is
well-described by uniform expansion at a rate of $0.642 \pm 0.049$\%
yr$^{-1}$ between 2007 and 2009, consistent with earlier observations.
Undecelerated expansion would imply an age of $156 \pm 11$ yr, but
reasonable expectations for either a Type Ia or core-collapse event
suggest deceleration with $R \propto t^{0.7}$, giving an age of about
110 yr or an explosion date of 1900.

\item The total X-ray flux is increasing at the rate of $1.7\pm
  1.0$\% yr$^{-1}$ (90\% confidence limits), a rate comparable to
  radio rates.  This increase makes it highly unlikely that G1.9+0.3
  is expanding into a stellar wind.

\item Direct determination of deceleration may be possible within the
lifetime of {\sl Chandra}.

\end{enumerate}

\acknowledgments

This work was supported by NASA through Chandra General Observer
Program grant G09-0062X.  AKC would like to thank E. Carlson for
discussions and helpful comments on the manuscript.

\newpage

\begin{deluxetable}{llc}

\tablecolumns{3}
\tablewidth{0pc}
\tabletypesize{\footnotesize}
\tablecaption{Best-Fit Parameters}

\tablehead{
\colhead{Parameter} & & Value
}

\startdata
Expansion Rate && (0.642$\pm$0.049)\% yr$^{-1}$ \\ 
Surface-Brightness Factor && $1.01 \pm 0.02$ \\ 
Flux Increase Rate (image)&& (1.89$\pm$0.86)\% yr$^{-1}$\\ 
Flux Increase Rate (spectrum)&&$(1.7 \pm 1.0)$\% yr$^{-1}$\\ 
Expansion Center   &RA   & 17:48:45.570$\pm 0.005$ \\ 
\qquad (J2000)     &Dec  &-27:10:06.94$\pm 0.08$ \\ 
\enddata
\tablecomments{Uncertainties are 90\% confidence intervals.
Errors on the expansion center are statistical only; {\sl Chandra}
pointing uncertainties are greater.
}

\end{deluxetable}

\begin{figure}
\epsscale{1.2}
\plotone{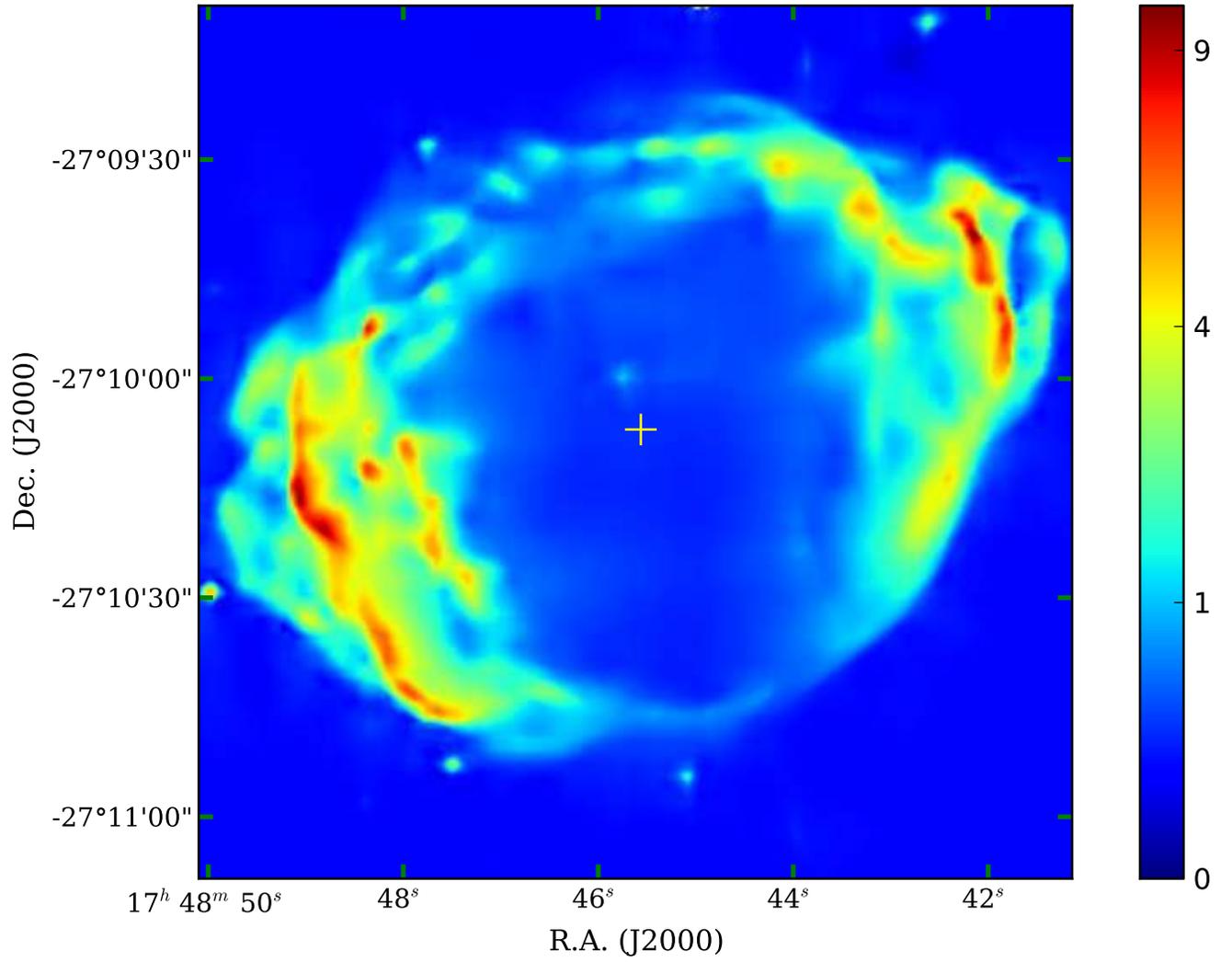}
\caption{{\sl Chandra} X-ray image from 2009,
smoothed with platelets \citep{willett07}.  The fitted expansion center
is indicated by the $+$ sign.  Surface brightness is in cts px$^{-1}$.
}
\end{figure}

\begin{figure}
\epsscale{1.0}
\plotone{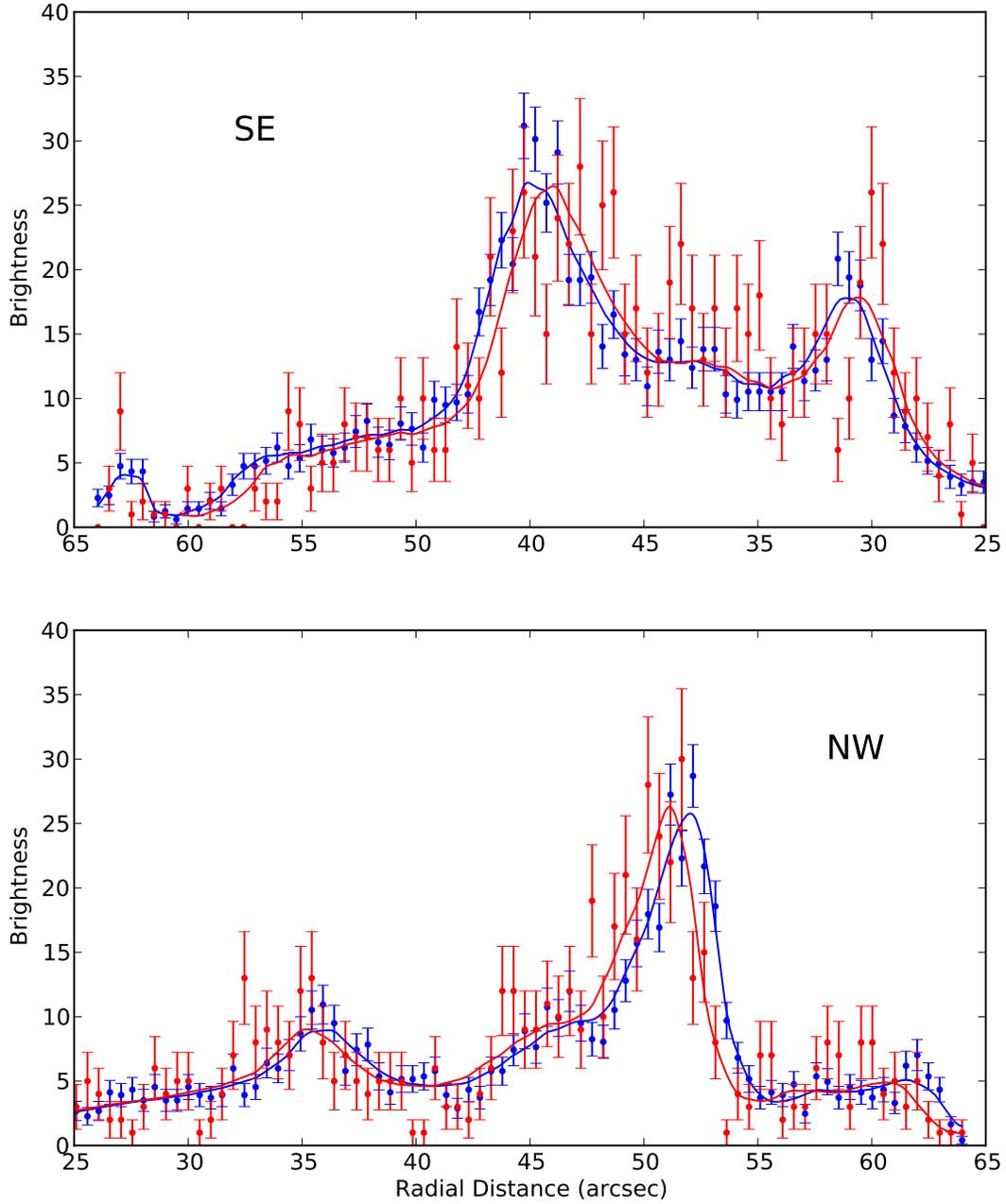}
\caption{Profiles from 2007 (red) and 2009 (blue) along position shown
in Figure 3. Horizontal scales are distance from the expansion center
in arcseconds. Top: SE limb; bottom, NW limb. Expansion is evident in
each.}
\label{profiles}
\end{figure}

\begin{figure}
\epsscale{0.9}
\plotone{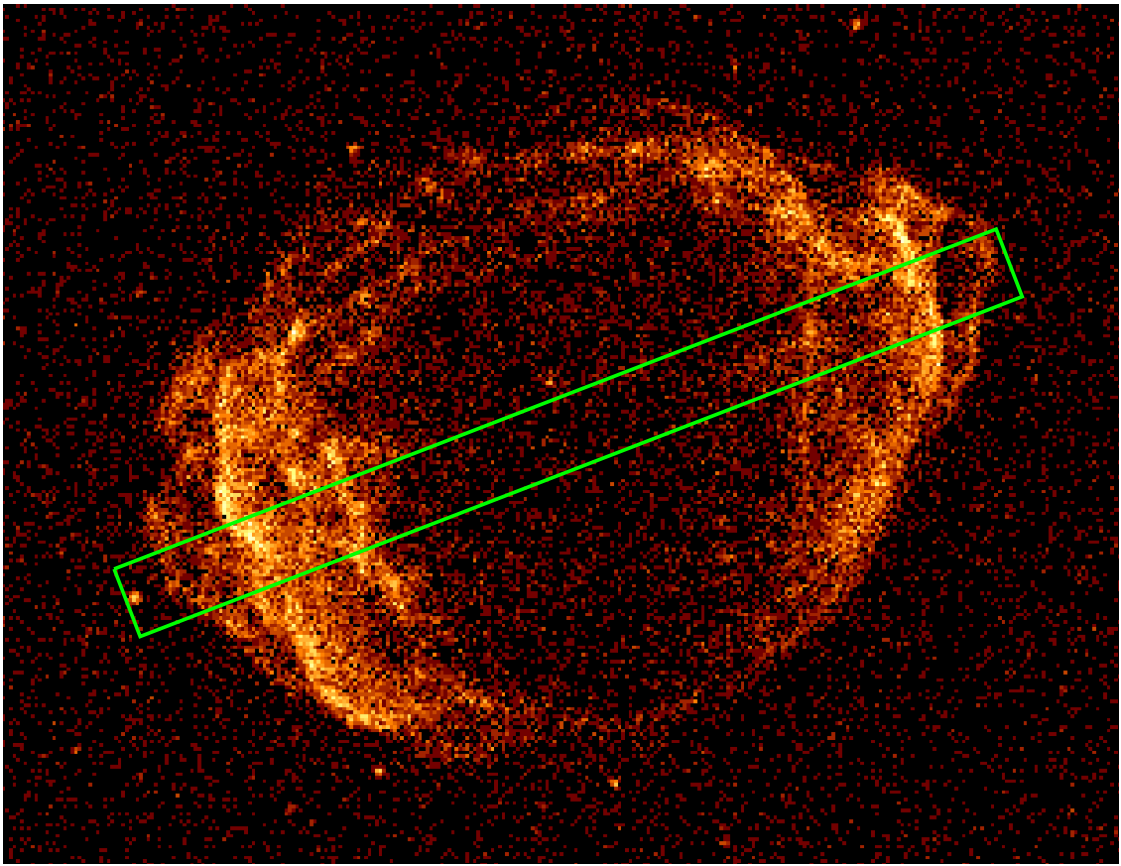}
\caption{Raw 2009 {\sl Chandra} image, showing region from which the
profiles of Fig.~2 were taken.}
\end{figure}

\end{document}